\documentclass[10pt]{article}
\usepackage{graphicx}
\usepackage{romannum}

\def\Title#1{\begin{center} {\Large #1 } \end{center}}
\def\Author#1{\begin{center}{ \sc #1} \end{center}}
\def\Address#1{\begin{center}{ \it #1} \end{center}}

\newcommand\pubblock{\rightline{\begin{tabular}{l} Proceedings of the Fifth Annual LHCP\\ \pubnumber\\
         \pubdate  \end{tabular}}}

\newenvironment{Abstract}{\begin{quotation} \begin{center} 
             \large ABSTRACT \end{center}\bigskip 
      \begin{center}\begin{large}}{\end{large}\end{center} \end{quotation}}

\newenvironment{Presented}{\begin{quotation} \begin{center} 
             PRESENTED AT\end{center}\bigskip 
      \begin{center}\begin{large}}{\end{large}\end{center} \end{quotation}}

\def\beq{\begin{equation}}
\def\eeq#1{\label{#1}\end{equation}}
\def\eeqn{\end{equation}}

\def\beqa{\begin{eqnarray}}
\def\eeqa#1{\label{#1}\end{eqnarray}}
\def\eeqan{\end{eqnarray}}

\let\bar=\overbar

\def\Dslash{\not{\hbox{\kern-4pt $D$}}}
\def\dslash{\not{\hbox{\kern-2pt $\del$}}}

\def\msb{{\bar{\ssstyle M \kern -1pt S}}}

\usepackage{color}

\newcommand\pubnumber{ CMS-CR-2017-194 }

\newcommand\pubdate{\today}

\def\affiliation{
On behalf of the CMS Collaboration, \\
Physics Institute \Romannum{3}A\\
RWTH Aachen University, Aachen, Germany}

\begin{document}

\large
\begin{titlepage}
\pubblock

\vfill
\Title{  Data Scouting : A New Trigger Paradigm }
\vfill

\Author{ Swagata Mukherjee }
\Address{\affiliation}
\vfill
\begin{Abstract}
In the year 2011, the CMS collaboration introduced the novel concept of data scouting, allowing to take data that otherwise would be rejected by the usual trigger filters. This special data flow, based on event-size reduction, was created to maintain sensitivity to new light resonances decaying to jets or muons, with very small online and offline resources allocated to it. The challenges implied by this new workflow and the solutions developed within the CMS experiment are shown. This technique is now a standard ingredient for CMS data-taking strategy. The present status of data scouting in CMS is presented. 
\end{Abstract}
\vfill

\begin{Presented}
The Fifth Annual Conference\\
 on Large Hadron Collider Physics \\
Shanghai Jiao Tong University, Shanghai, China\\ 
May 15-20, 2017
\end{Presented}
\vfill
\end{titlepage}
\def\thefootnote{\fnsymbol{footnote}}
\setcounter{footnote}{0}
%

\normalsize 


\section{Introduction}
The Compact Muon Solenoid (CMS) \cite{Chatrchyan:2008aa} is a general purpose detector designed for the precision
measurement of leptons, photons, and jets and other physics objects in proton-proton collisions at the CERN LHC. 
The LHC is designed to collide protons at a center-of-mass energy of 14 TeV and a luminosity of $10^{34}$ $\mathrm{cm}^{-2}\mathrm{sec}^{-1}$. 
At design luminosity, the $pp$ interaction rate can exceed 1 GHz.
The digital readout of CMS detector generates huge amount of raw data, much of which can be useful for physics analysis. 
However, full reconstruction of all collision events is not feasible with existing computing resources. 
CMS uses a two-level trigger strategy \cite{Khachatryan:2016bia}, consisting of a Level-1 Trigger (L1) and a High-Level Trigger (HLT), to reduce the data volume to 
approximately 1 kHz of fully reconstructed physics events. 
The CMS L1 Trigger is implemented in custom hardware and performs a preliminary selection on physics events, reducing the 1 GHz rate of input events to approximately 100 kHz. The HLT is a software application, consisting of more than 500 trigger paths, each selecting for a particular physics signature. Approximately 1 kHz of events are selected by the HLT and sent to the prompt and offline reconstruction system.
To make accept or reject decision for each event, the HLT performs physics object reconstruction and applies a selection based on the characteristics of the reconstructed objects. Trigger paths consist of sequences of producer modules, which build collections of objects; and filter modules, which reject events that do not fulfill certain criteria. Most trigger paths contain multiple phases of reconstruction and filtering, and generally the later phases of reconstruction yield physics objects whose performance is quite close to that of their offline counterparts. If an event reaches the very end of any trigger path, it is accepted. The factors that restrict the rates that the 
HLT can achieve are: the amount of storage space and the maximum throughput of the data acquisition system (DAQ), the capacity of 
the offline prompt reconstruction system. 

In this note we describe data scouting \cite{Anderson:2016ron}, a technique that relies on the online reconstruction of physics objects in order to attain 
fairly high trigger rates. Scouting provides new opportunities for physics analysis by overcoming the usual limitations of the traditional trigger strategy.

\section{Data Scouting Technique}
In the regime of data scouting, events are recorded at the highest possible rate while providing reconstructed objects whose performance is suitable for a physics analysis performed offline. To do this, the online reconstruction algorithms available at the HLT is used. HLT physics objects are less performant than their offline counterparts, but for many analyses, the difference does not significantly affect the sensitivity. Saving the objects reconstructed at the HLT, instead of those reconstructed offline, is less resource-consuming. This online reconstruction strategy is implemented via data scouting streams at the HLT. Each data scouting stream contains a number of trigger paths (scouting triggers), which perform event reconstruction and selection in the same way that ordinary HLT paths do. However, the selection criteria are much looser than for ordinary paths, and hence the rate of trigger firing is higher.
For events passing one or more scouting triggers, additional online reconstruction sequences are run in order to produce all physics objects necessary for an offline measurement or search. The produced objects are converted to a special compact event format and saved to disk. The data recorded by the scouting triggers is made available offline and can be used for physics analysis. For these events, no offline reconstruction is performed, and the raw data is not saved.
The reduced, compact event format of scouting requires negligible space on disk and does not place any 
additional strain on the DAQ system. Size of the scouting events are 100 to 1000 times smaller than the standard raw data format. 
Scouting has been used to increase the total number of CMS events recorded for physics by a factor of 2-6 beyond what the standard trigger strategy provides.

\section{Data Scouting in LHC Run \Romannum{1}}
The first scouting triggers were tested in CMS during the last few $pp$ fills of the 2011 LHC run. 
The scouting triggers successfully collected 0.13 fb$^{-1}$ data. Events with $H_{\mathrm{T}}$ (defined as the scalar sum of jet transverse momenta) larger than 350 GeV were recorded and saved in a reduced format, containing only the set of jets reconstructed from particle-flow (PF) candidates \cite{Sirunyan:2017ulk} by the HLT algorithm. The data were used to perform a search for heavy resonances decaying to dijets \cite{CMS:2012cza}. The search demonstrated sensitivity to resonances with masses between 0.6 and 0.9 TeV, a parameter region inaccessible to the standard CMS dijet resonance search. 
\begin{figure}[htb]
\centering
\includegraphics[height=2in]{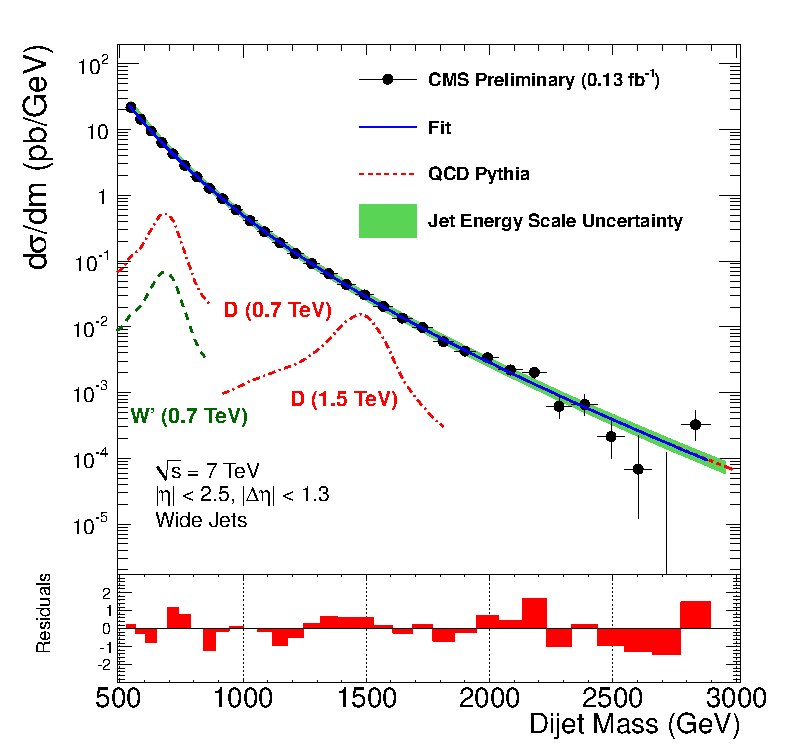}
\includegraphics[height=2in]{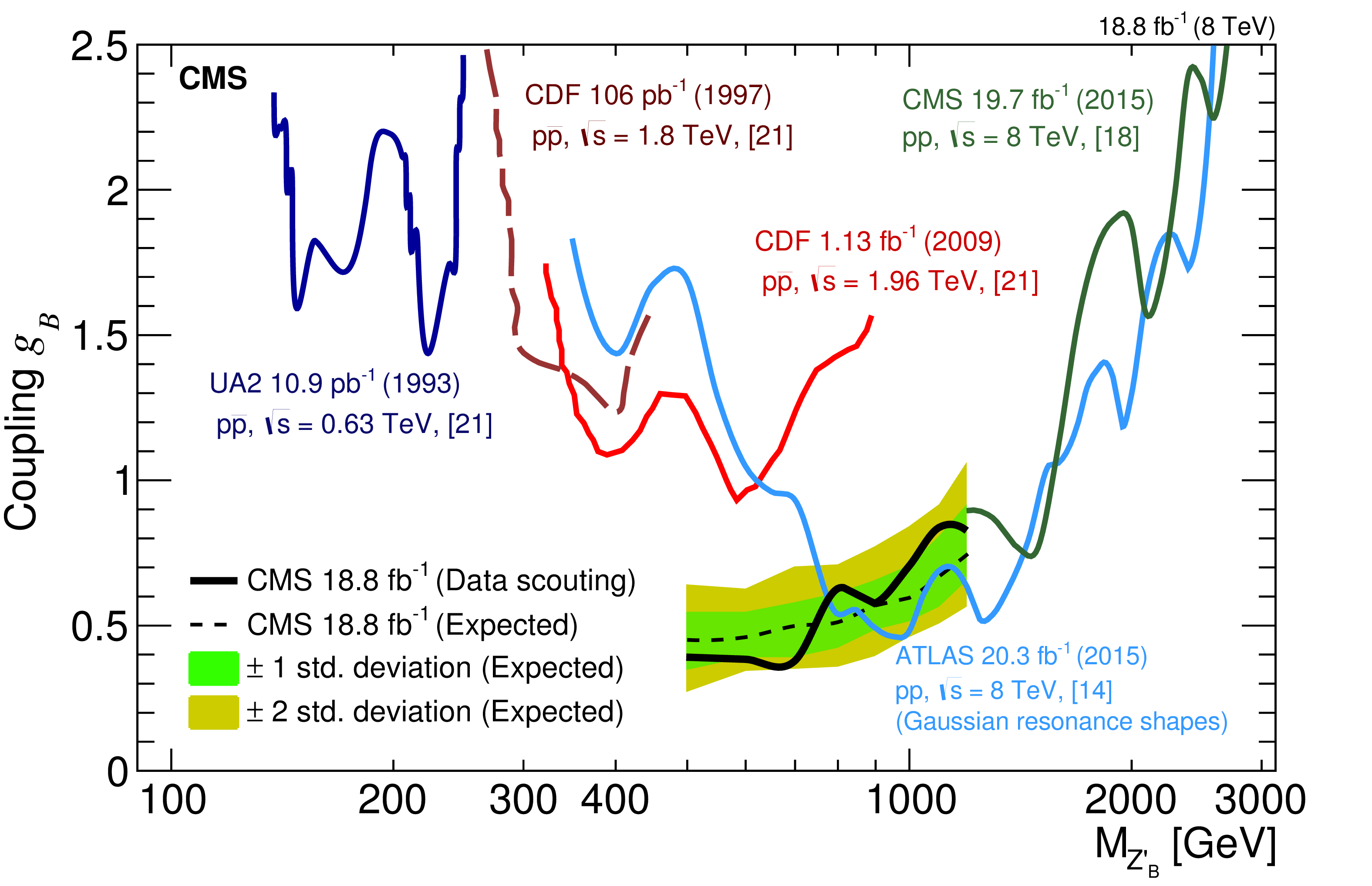}
\caption{ Left: Dijet mass spectrum from HLT wide jets (points) compared to a smooth fit (solid)
and to predictions including simulation of QCD (short-dashed), di-quark signals
(dot-dashed), and a $W'$ signal (long-dashed). The data was taken in 2011 at $\sqrt{s}=7$ TeV~\cite{CMS:2012cza}. The QCD prediction has been normalized to the
data. The error bars are statistical only. The shaded band shows the contribution from
the systematic uncertainty in the jet energy scale (JES). The bin-by-bin fit residuals are shown
at the bottom. Right: Observed 95\% CL upper limits on the coupling $g_B$ of a hypothetical leptophobic resonance $Z'_B \rightarrow qq'$ as a function of its mass with the data taken in 2012 
at $\sqrt{s}=8$ TeV~\cite{Khachatryan:2016ecr}. The results from this study are compared to results obtained with similar searches  at different collider energies.}
\label{fig:figure1}
\end{figure}

After the 2011 test demonstrated the viability of the scouting technique, the strategy was repeated for the full 2012 CMS dataset. The scouting trigger selection cut was lowered to 
$H_{\mathrm{T}}>250$ GeV to accommodate an even larger rate of events. Due to CPU concerns related to the high rate, calorimeter jets were reconstructed and saved instead of PF jets. The collected data, corresponding to 18.8 fb$^{-1}$, were used to perform a dijet resonance search \cite{Khachatryan:2016ecr} analogous to the one carried out in 2011. The search results were interpreted as limits on the mass and coupling of a hypothetical leptophobic $Z'$ resonance decaying to quarks. As indicated in Fig.\ref{fig:figure1}-right, the limits were the strongest that were obtained for masses between 0.5 and 0.8 TeV, improving markedly on results from previous colliders.

\section{Data Scouting in LHC Run \Romannum{2}}
The success of data scouting in LHC Run I leaded to an expansion of the strategy for Run II.
The aim was to maintain the ability to search very low in $H_{\mathrm{T}}$ using calorimeter jets, as already done in 2012,
while also providing an event format capable of supporting a broader range of scouting analyses.
Two streams were deployed at the HLT for data taking in 2015 and 2016: one saving an event
content based on calorimeter objects (the calo-scouting stream) 
and one saving an event content based on PF objects (the PF-scouting stream).
Each stream had its own set of trigger paths and output datasets. 
Table 1 summarizes the rates and bandwidths measured for each stream in 2016.

\begin{table}[t]
\begin{center}
\begin{tabular}{ccc}  
Scouting trigger path    &    Rate at 10$^{34}$ cm$^{-2}$ sec$^{-1}$  &  Bandwidth (MB/s) \\ \hline
Calo scouting $H_{\mathrm{T}}$    &    3700                                                     &    11  \\
PF scouting $H_{\mathrm{T}}$       &    720                                                       &    9  \\
PF scouting DiMuon      &   480                                                        &    6  \\
Commissioning              &    30                                                         &    $<$1 \\
Monitoring                       &    26                                                         &   23 \\ \hline
\end{tabular}
\caption{Trigger rates and bandwidths for various scouting triggers measured in 2016 data.}
\label{tab:table1}
\end{center}
\end{table}

Triggers in the calo-scouting stream reconstruct jets from calorimeter deposits. The main
signal trigger in this stream selects events with $H_{\mathrm{T}}>250$ GeV. 
The event content for this stream includes the reconstructed calorimeter jets, the missing transverse momentum (MET), 
and $\rho$,  a measure of the average energy density in the event. Local pixel track reconstruction provides b-tagging information for the jets.
The size of this event content is about 1.5 kB on average for events passing the $H_{\mathrm{T}}>250$ GeV trigger.

Triggers in the PF-scouting stream run the online version of the full PF sequence to reconstruct
selected events. The main signal trigger in this stream selects events with $H_{\mathrm{T}}>410$ GeV. 
Additionally, the stream 
contains a trigger path selecting events with two muons having invariant mass above 3 GeV; this
represents an effort to extend scouting to searches in final states other than fully-hadronic ones.
The event content for this stream includes the reconstructed PF jets, the PF MET, $\rho$, a collection 
of primary vertices, and all PF candidates with $p_{\mathrm{T}}>0.6$ GeV. It also contains electron, muon, and photon objects. The size of this event content 
is approximately 10 kB per event. Auxiliary prescaled trigger paths
are also included both in calo and PF streams in order to facilitate measurements of the signal trigger efficiency.

To facilitate comparisons of the online physics objects to their offline reconstructed counterparts, and to monitor the quality of the scouting data, a separate monitoring stream is deployed. This stream contains prescaled versions of all scouting triggers in the calo-scouting and PF-scouting streams. Events selected for this stream are saved in reduced scouting event format, but they are also sent to the CMS prompt reconstruction system for offline processing. This stream therefore contains online and offline reconstructed versions of each event, 
which enables a detailed object-by-object comparison of the online and offline reconstruction performance.

A drawback of the scouting approach is discarding of the raw data. In case of a hint of a discovery in scouting data, it is useful to have the full raw data available so that the events can be analyzed in greater detail to scrutinise all aspects of the possible signal. This is accomplished using parked triggers, which send selected events directly to storage disk without offline reconstruction.
\begin{figure}[htb]
\centering
\includegraphics[height=2.8in]{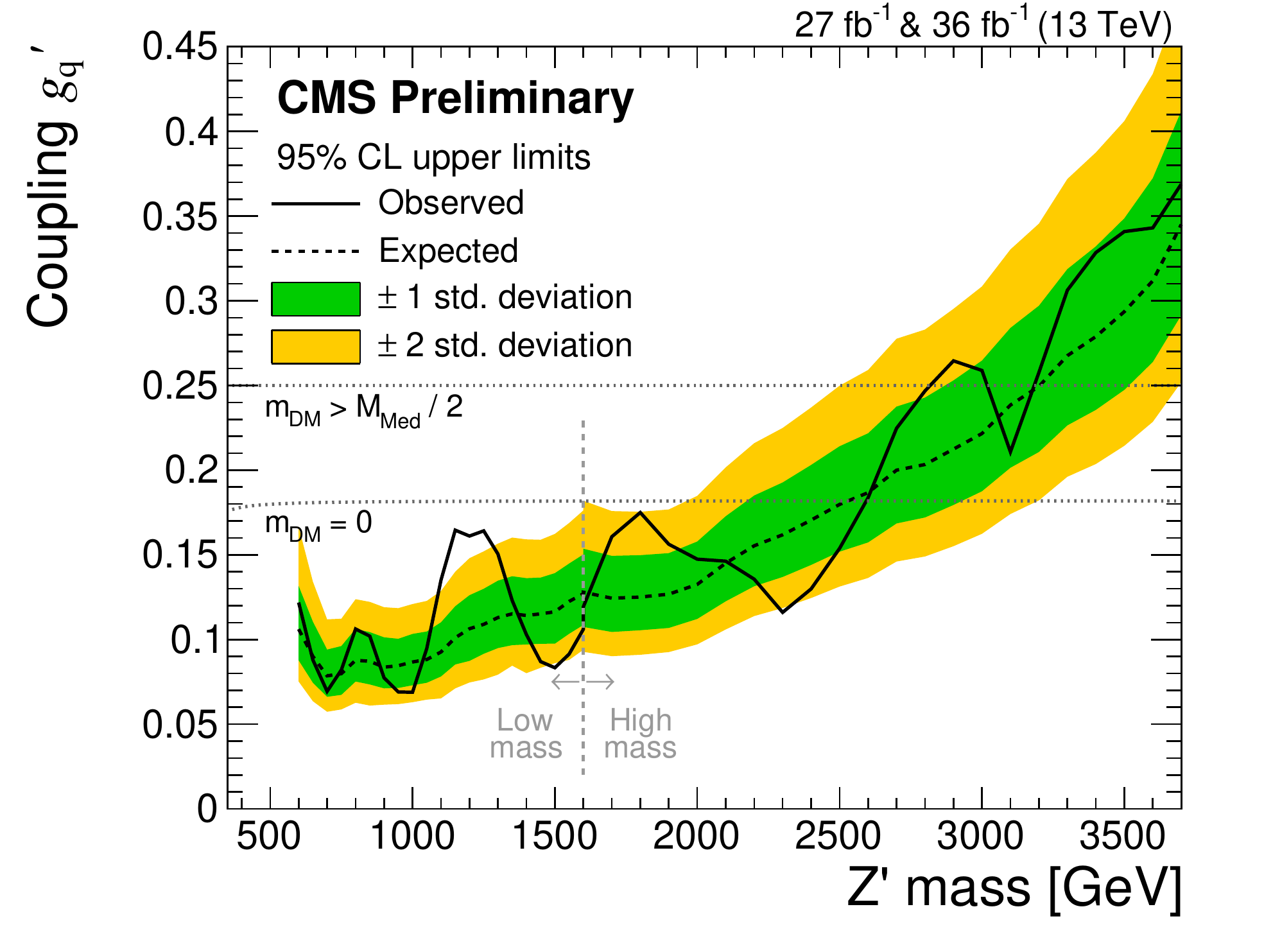}
\caption{ The 95\% CL upper limits on the universal quark coupling $g_{q'}$ as a function of resonance mass for a leptophobic $Z'$
resonance that only couples to quarks. The observed limits (solid), expected limits (dashed) and their variation at the 1 and 2 standard deviation levels (green and yellow shaded bands) are shown~\cite{CMS:2017xrr}.}
\label{fig:figure2}
\end{figure}

Data collected in the calo-scouting stream in 2016 were used to perform a search for dijet
resonances with masses between 0.6 TeV and 1.6 TeV \cite{CMS:2017xrr}. The search was carried out on 
27 fb$^{-1}$ of $pp$ collision data. 
Fig.\ref{fig:figure2} shows upper limits on the coupling as a function of mass for a model of a leptophobic $Z'$ resonance with a 
universal quark coupling, $g_{q'}$ \cite{Boveia:2016mrp}, related to the Z' coupling convention of \cite{Dobrescu:2013coa} by $g_{q'}=g_{B}/6$.


\section{Conclusions}
The novel data scouting technique allows events to be collected at a rate much higher than what is normally achievable with the CMS trigger system. It is implemented 
in a way so that no major change is needed to the basic HLT infrastructure and it does not place a strain on the DAQ system, disk resources, or the reconstruction system.
The two scouting streams, calo and PF, deployed for Run II of the LHC opens up various possibilities of search of New Physics. The calo stream is oriented towards dijet resonance searches and the PF stream is designed to support arbitrary searches for a variety of final states. The first physics results using the Run II scouting framework has been made public. The CMS collaboration will take full advantage of this tool to search in other regions, which was previously unexplored at the LHC.


\end{document}